\documentclass[pdflatex,sn-nature]{sn-jnl}% Style for submissions to Nature Portfolio journals

\usepackage{graphicx}%
\usepackage{multirow}%
\usepackage{amsmath,amssymb,amsfonts}%
\usepackage{xcolor}%
\usepackage{textcomp}%
\usepackage{manyfoot}%
\usepackage{listings}%
\usepackage[switch]{lineno}
\usepackage{txfonts}[T1]
\usepackage{booktabs}

\begin{document}

\title[Evidence of star cluster migration and merger in dwarf galaxies]{Evidence of star cluster migration and merger in dwarf galaxies}

\author*[1]{\fnm{Mélina} \sur{Poulain}}\email{melina.poulain@oulu.fi}\email{melina.poulain45@gmail.com}

\author[2]{\fnm{Rory} \sur{Smith}}

\author[3]{\fnm{Pierre-Alain} \sur{Duc}}%\email{iiiauthor@gmail.com}

\author[4]{\fnm{Francine R.} \sur{Marleau}}%\email{iiauthor@gmail.com}

\author[5]{\fnm{Rebecca} \sur{Habas}}

\author[6]{\fnm{Patrick R.} \sur{Durrell}}

\author[7]{\fnm{Jérémy} \sur{Fensch}}

\author[8]{\fnm{Sungsoon} \sur{Lim}}

\author[9,10,11]{\fnm{Oliver} \sur{Müller}}

\author[12]{\fnm{Sanjaya} \sur{Paudel}}

\author[13]{\fnm{Rubén} \sur{Sánchez-Janssen}}

\affil*[1]{Space Physics and Astronomy Research Unit, University of Oulu, P.O. Box 3000, FI-90014, Oulu, Finland}

\affil[2]{Universidad Técnica Federico Santa María, Avda. Vicuña Mackenna 3939, San Joaquín, Santiago, Chile}

\affil[3]{Universit\'e de Strasbourg, CNRS, Observatoire astronomique de Strasbourg, UMR 7550, F-67000 Strasbourg, France}

\affil[4]{Universität Innsbruck, Institut für Astro- und Teilchenphysik, Technikerstraße 25/8, 6020 Innsbruck, Austria}

\affil[5]{INAF – Astronomical Observatory of Abruzzo, Via Maggini, 64100 Teramo, Italy}

\affil[6]{Dept. of Physics, Astronomy, Geology, and Environmental Sciences, Youngstown State University, Youngstown, OH 44555 USA}

\affil[7]{Univ. Lyon, ENS de Lyon, Univ. Lyon 1, CNRS, Centre de Recherche Astrophysique de Lyon, UMR5574, 69007 Lyon, France}

\affil[8]{Department of Astronomy, Yonsei University, 50 Yonsei-ro Seodaemun-gu, Seoul, 03722, Republic of Korea}

\affil[9]{Institute of Physics, Laboratory of Astrophysics, École Polytechnique Fédérale de Lausanne (EPFL), 1290 Sauverny, Switzerland}

\affil[10]{Institute of Astronomy, Madingley Rd, Cambridge CB3 0HA, UK}

\affil[11]{Visiting Fellow, Clare Hall, University of Cambridge, Cambridge, UK}

\affil[12]{Department of Astronomy and Center for Galaxy Evolution Research, Yonsei University, Seoul 03722}

\affil[13]{UK Astronomy Technology Centre, Royal Observatory Edinburgh, Blackford Hill, Edinburgh EH9 3HJ, UK}


%max 200 words
%Articles start with a fully referenced summary paragraph, ideally of no more than 200 words, which is separate from the main text and avoids numbers, abbreviations, acronyms or measurements unless essential. It is aimed at readers outside the discipline. This summary paragraph should be structured as follows: 2-3 sentences of basic-level introduction to the field; a brief account of the background and rationale of the work; a statement of the main conclusions (introduced by the phrase 'Here we show' or its equivalent); and finally, 2-3 sentences putting the main findings into general context so it is clear how the results described in the paper have moved the field forwards.
%Example: https://www.nature.com/documents/nature-summary-paragraph.pdf

\abstract{
%2-3 sentences of basic-level introduction to the field
Nuclear star clusters (NSCs) are the densest stellar systems in the Universe. They can be found at the center of all galaxy types, but tend to favor galaxies of intermediate stellar mass around 10$^9$\,M$_{\odot}$\cite{Cote2006,Janssen2019}. Currently, two main processes are under debate to explain their formation: in-situ star-formation from gas infall\cite{Loose1982} and migration and merging of globular clusters (GCs) caused by dynamical friction\cite{Tremaine1975}.
%Two to three sentences of more detailed background, comprehensible to scientists in related disciplines.
Studies\cite{Turner2012,Johnston2020,Fahrion2021,Fahrion2022,Fahrion2022b} of NSC stellar populations suggest that the former predominates in massive galaxies, the latter prevails in dwarf galaxies, and both contribute equally at intermediate mass. 
%One sentence clearly stating the general problem being addressed by this particular study
However, up to now, no ongoing merger of GCs has yet been observed to confirm this scenario.
%a statement of the main conclusions (introduced by the phrase 'Here we show' or its equivalent)
Here we report the serendipitous discovery of five dwarf galaxies with complex nuclear regions, characterized by multiple nuclei and tidal tails, using high resolution images from the Hubble Space Telescope. These structures have been reproduced in complementary N-body simulations, supporting the interpretation that they result from migrating and merging of star clusters.
%2-3 sentences putting the main findings into general context so it is clear how the results described in the paper have moved the field forwards.
The small detection rate and short simulated timescales (below 100\,Myr) of this process may explain why this has not been observed previously. This study highlights the need of large surveys with high resolution to fully map the migration scenario steps.
}

%\keywords{keyword1, Keyword2, Keyword3, Keyword4}



\maketitle
\twocolumn
%\linenumbers

\renewcommand{\figurename}{\footnotesize Fig.}

Witnessing the formation of nuclear star clusters (NSCs) is important to fully understand how such extremely compact and massive objects can form at the center of a wide range of galaxy type. NSCs typically have a stellar mass M$_*$ in the range $10^5$ to $10^8$ M$_{\odot}$ and an effective radius R$_e$ up to several tens of parsecs\cite{Neumayer2020}. 
Over the last decades, two main scenarios have been put forward and are debated to explain the formation of NSCs.

In the in-situ formation scenario\cite{Loose1982}, the infall of gas triggers star formation and forms a NSC. This scenario is favored in massive galaxies and predicts the presence of two compact sources: one is either a massive black hole or a star cluster surrounded by a stellar disk, while the other compact object is being formed from stars gathering at the disk apoapsis. This is observed in the nuclear region of the Andromeda galaxy\cite{Lauer1993}, and has been reported in several other galaxies\cite{Lauer1996,Debattista2006,Menezes2018}.

Alternatively, NSCs could form from the infall of GCs due to dynamical friction\cite{Tremaine1975}. This migration plus merging scenario is suggested to dominate in dwarf galaxies. Observational signatures of this scenario include the presence of multiple star clusters\cite{Georgiev2014,Schiavi2021,Fahrion2024} and tidal interactions in the inner regions, possibly leading to the formation of tails. However, no direct observation of tidally interacting and merging star clusters near the center of dwarf galaxies have been reported so far in the literature.

The study of the stellar population alone is often not enough to disentangle between both scenarios. Young stellar populations, in particular, can result from either formation scenario, or a combination of the two. An example of this combination is the wet-merger scenario\cite{Guillard2016}, i.e the formation of a NSC from the migration of a star cluster together with a gas reservoir. Thus, such studies need to be coupled to high resolution observations to fully reconstruct the formation steps of NSCs.

Involving interacting galaxies, a NSC growth scenario is possible. In the stages of a merger between two nucleated galaxies, multiple nuclei should be visible in the central regions of the galaxies, as observed in several systems\cite{Pak2016}. The nuclei will eventually end up merging to form the NSC of the galaxy remnant. A clear sign of ongoing or past galactic collisions is the presence of an overall boxy shape of the galaxy remnant, or tidal tails and shell structures in its outskirts.

High resolution optical images of a sample of 79 dwarf galaxies were recently obtained with the Hubble Space telescope (HST) as part of follow-up observations of nearby galaxy satellites from the Mass Assembly of early-Type gaLAxies with their fine Structures (MATLAS) survey (see Methods). A vast majority of the selected galaxies for the HST follow-up have a lower surface brightness and a larger size than typical dwarfs, and can be defined as ultra-diffuse-galaxies\cite{Marleau2024} (UDGs). The HST sample covers about 65\% of the MATLAS UDG sample\cite{Marleau2021}, and apart from their size and surface brightness criteria, UDGs share similar structural properties to dwarfs. Among the galaxies observed with HST, 10 exhibit a nucleus with substructure, such as multiple star clusters and stellar tidal tails.
We note that bright sources are observed on the deep ground-based MATLAS images at the location of the complex nuclear systems revealed by the high resolution HST images (Extended Data Fig. 1 and 2). This excludes the possibility that these nuclear substructures are instrumental artifacts.

\begin{figure}
\centering
\includegraphics[width=\linewidth]{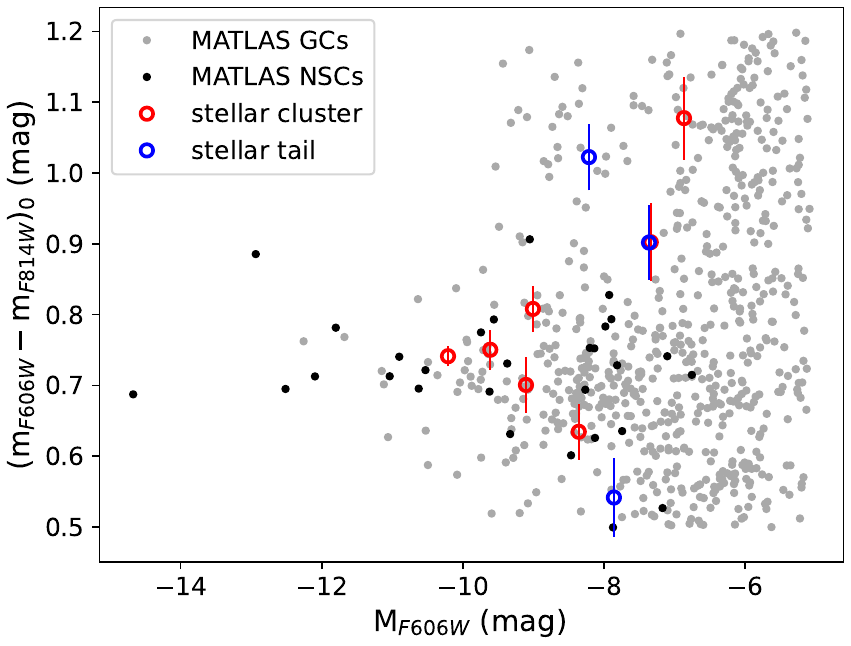}
\caption{\footnotesize \textbf{Color-magnitude diagram of the substructures of the five nuclear regions consistent with a migration scenario origin.} We compare the star clusters (\emph{red circles}) and tails (\emph{blue circles}) of their complex nuclear regions to GCs (\emph{grey dots}) and NSCs (\emph{black dots}) of the MATLAS dwarfs observed by HST.}
\label{fig:color-mag}
\end{figure}

To check whether the observed nuclear substructures are signatures of the migration scenario, i.e., the predicted formation process for the nucleus of dwarf galaxies, we isolated a sample of dwarfs whose nuclear structures likely formed from internal mechanisms and whose luminosity and color are consistent with NSCs and GCs from the MATLAS survey (details in Methods). Therefore, we removed four dwarf-dwarf merger candidates (Extended Data Fig. 2), and excluded one NSC that might be experiencing a wet-merger scenario given the blue color of some structures of its nucleus (Extended Data Table \ref{tab:photometry}). Hence, we are left with a sample of five galaxies. A color-magnitude diagram of their substructures is shown in Figure \ref{fig:color-mag}.

We computed the detection rate of interacting/merging nuclei in our specific sample of dwarfs observed with the HST. Of the whole initial sample, only 13\% (10/79) show a complex nuclear region. Moreover, the fraction decreases to 7\% (5/74) when we exclude the dwarf-dwarf merger candidates, and to 4\% (3/74) if we focus only on the least ambiguous sign of a star cluster merger, tidal tails. This detection rate might be higher than that of typical dwarfs because the low central surface brightness of our galaxies makes easier the detection of tidal tails. Thus, observing complex nuclei, and especially tidal tails, is rather rare and requires a large sample of low surface brightness dwarfs.

\begin{figure*}
\centering
\includegraphics[width=\textwidth]{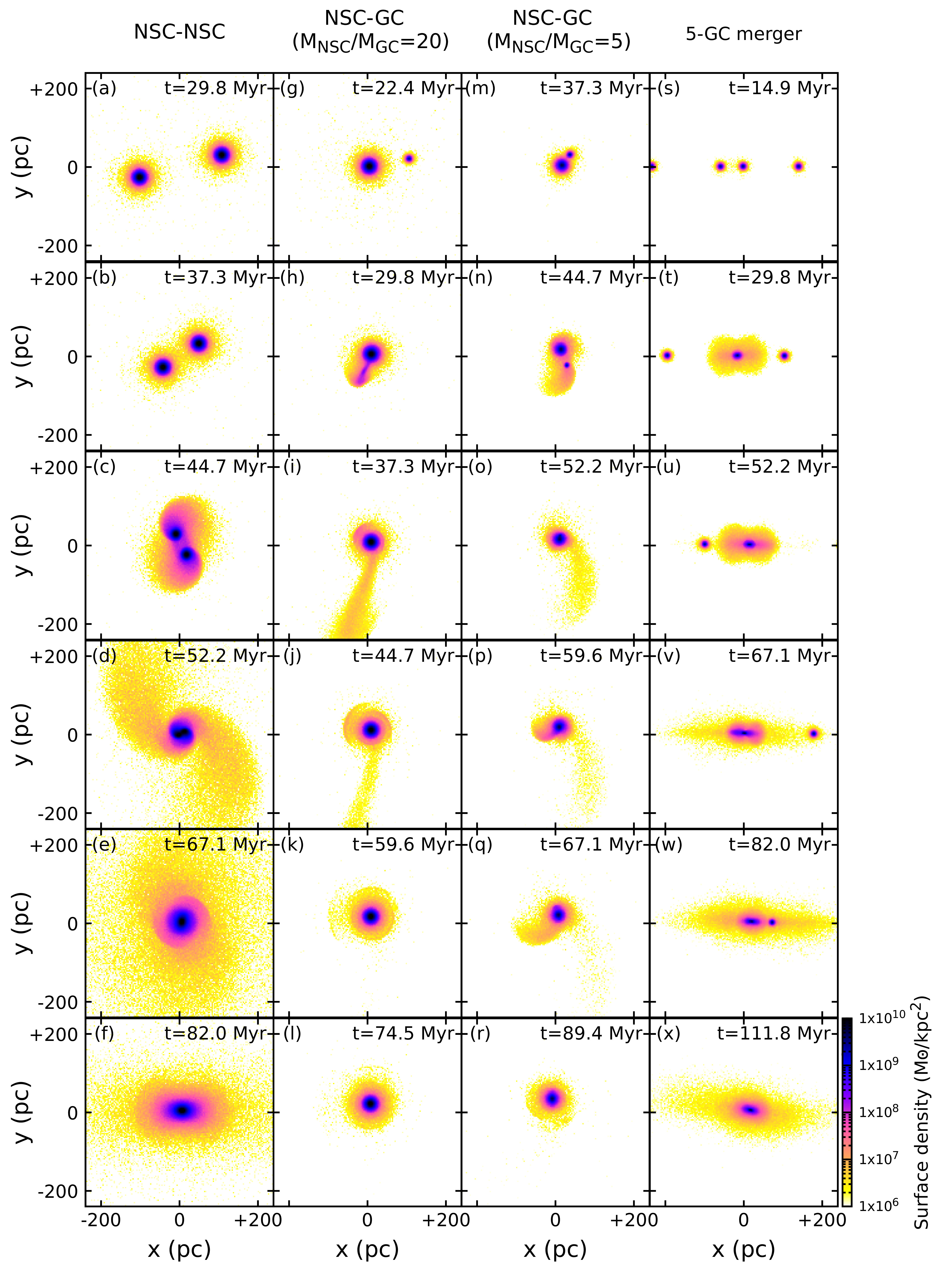}
\caption{\footnotesize \textbf{Snapshots illustrating the key steps of the N-body simulations of merging compact stellar objects.} From \emph{(a)} to \emph{(f)}: a NSC-NSC merger with non-null tangential velocity. From \emph{(g)} to \emph{(l)}: a NSC-GC merger with a mass ratio 1:20. From \emph{(m)} to \emph{(r)}: an 1:5 NSC-GC merger. From \emph{(s)} to \emph{(x)}: a 5-GC merger. Assuming a stellar mass-to-light ratio of 7, a surface density of 3.1(1.2)$\times$10$^6$~M$_\odot$/kpc$^2$ translates in a surface brightness of 27(28) mag\,arcsec$^{-2}$ (\emph{r}-band filter).}
\label{fig:sim}
\end{figure*}

To investigate the origin of the nuclear stellar tails, we compared our observations with the results of collisionless N-body simulations of NSC-NSC, NSC-GC, and GC-GC mergers. We explored the effect of the simulation parameters, e.g., the impact parameters, radial velocity, tangential velocity, or mass ratio, on the merging process. Based on the average properties of NSCs in the MATLAS dwarfs observed with HST\cite{Poulain2024} and the properties of the Milky Way and the Andromeda galaxy GCs\cite{Harris1996,Peacock2010}, we used an initial stellar mass M$_*=10^{6.5}$ and $10^{5.2}$\,M$_{\odot}$ for NSCs and GCs, respectively. Further details on the simulations can be found in the Methods section. In Figure \ref{fig:sim}, we show the main observed effects of these parameters on the different types of simulations. 
Our results suggest that overall, the merger of two star clusters occurs on short timescales, typically within a maximum of $50\ $Myr from first contact between the clusters to complete coalescence. We note the clusters must collide for the merger to happen on these short timescales. If the tangential or radial velocity is high enough so that they do not make contact, then they would orbit past each other without creating any features, and perhaps never merge. 

The merger of clusters of similar mass (1:1), i.e. two NSCs or two GCs, induces the formation of an extended elliptical nucleus without a long and extended tail, as represented in snapshots (\emph{a}) to (\emph{f}). When the merger involves more than two GCs, as shown in snapshots (\emph{s}) to (\emph{x}), the larger the number of clusters, the longer the extension of the final remnant nucleus. In addition, an S-shape structure can be visible for a few tens Myr when a non-null tangential velocity is applied to the colliding cluster, as observed in snapshots (\emph{c}) and (\emph{d}). 
 
A NSC-GC merger, or a merger of star clusters with mass ratios of at least 1:5 or greater, produces long and extended stellar tails together with shell structures around the newly formed NSC. A 1:20 and 1:5 merger are represented in snapshots (\emph{g})$-$(\emph{l}) and (\emph{m})$-$(\emph{r}), respectively. Assuming an old red stellar population in the clusters and a detection limit of 28\,mag\,arcsec$^{-2}$, tails are visible for about $30$ to $40\ $Myr. Shells at that surface brightness or brighter remain for a longer time ($45-90$ Myr). However, the detailed structures of the shells are more likely to appear diffuse at the resolution of HST, making them more difficult to identify in our observations.

The length of the tidal tail depends on the difference of mass between the two clusters, where the larger the mass ratio, the longer the tail. For example, the maximum length of the tail changes from about 480 pc for a 1:20 mass ratio to only 275 pc for a 1:5 mass ratio. Moreover, in the case of small mass ratios ($M_{\rm NSC}/M_{\rm GC} = 5$), we observe the apparition of a smaller secondary tail, noticeable in snapshots (\emph{p}) and (\emph{q}). 

We note that, a tangential velocity of 2\,km/s, or less, did not affect the overall morphology of the NSC-GC mergers. However, for a 1:20 ratio and a tangential velocity similar to the galaxy circular velocity (here 5\,km/s), multiple pericenter passages are necessary before the merger occurs, each of them producing a very short lasting ($\sim$7.5\,Myr) and small tail ($\sim$100\,pc at 28\,mag\,arcsec$^{-2}$), while the merger produces a longer tail (up to 250\,pc) visible for about 38 Myr at 28 mag\,arcsec$^{-2}$. 

We further investigated the effect of the stellar and dark-matter components of the host galaxy on the formation of the tidal tails. The use of a cuspy halo or more massive disk slightly shortens the timescale on which the merger happens compared to our fiducial dwarf galaxy model, due to increased dynamical friction. But, it has only a mild impact on the duration for which the streams will be visible. We observe that the tidal tails last up to 60 Myr at 28\,mag\,arcsec$^{-2}$ when using a ten times more massive stellar disk.
This suggests that the detection rate should increase with the stellar mass of the galaxy. However, we find that the tidal features, with their low surface brightness, will be more easily hidden by the intrinsic brightness of the higher surface brightness dwarfs, and thus be more difficult to detect in a typical dwarf than in UDGs. 
Overall, a change in the stellar or dark-matter components does not seem to have a huge impact on the general shape of the created tidal tails, and while the features will be longer detectable with the former, the timescales are still short, less than 100 Myr.

\begin{figure*}
\centering
\includegraphics[width=0.95\textwidth]{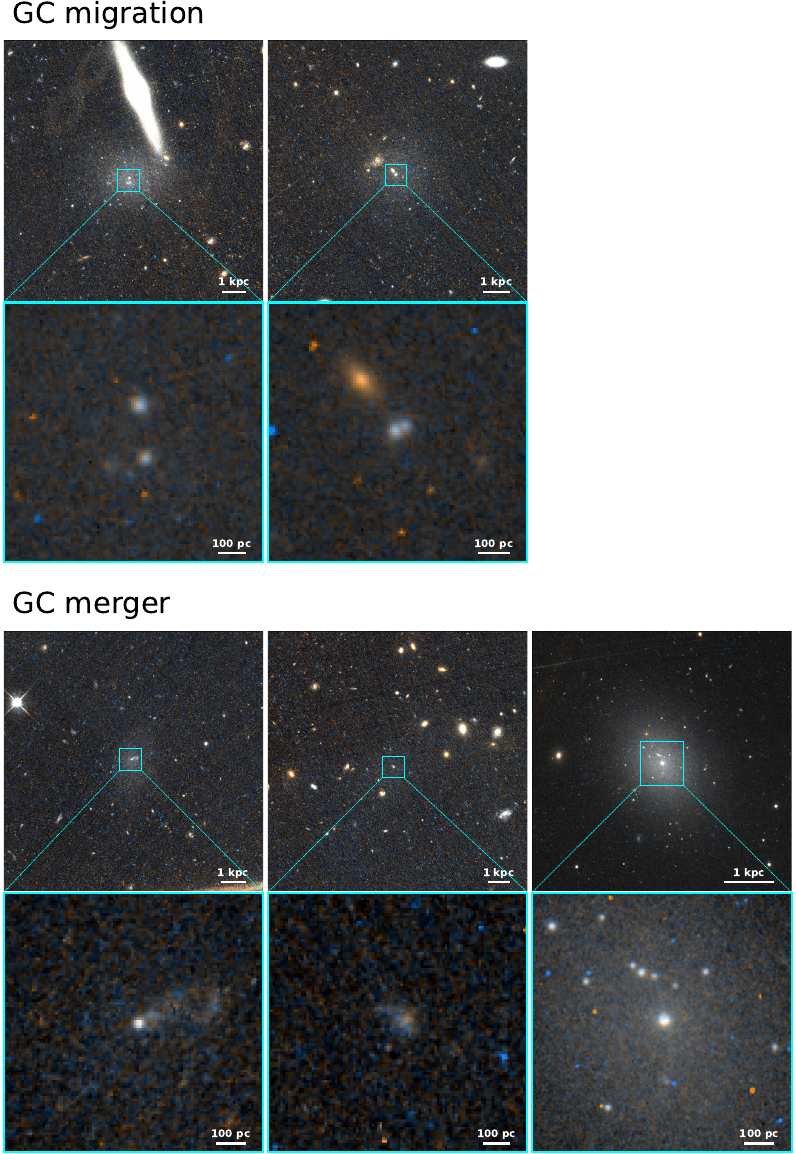}
\caption{\footnotesize \textbf{Mains steps of the migration and merging of GCs in dwarf galaxies.} The sequence starts with the migration of massive GCs towards the galaxy center, as seen in the two double nucleated dwarfs MATLAS-138 and MATLAS-987, followed by a GC-NSC merger that exhibit tidal tails, as in MATLAS-1216, MATLAS-207, MATLAS-1938. We show $1'\times1'$ and $5"\times5"$ HST F606W and F814W color composed cutouts of the galaxies and their nuclear region, respectively. The zoom on MATLAS-1938 nucleus is of $10"\times10"$ to highlight the elongation of the stellar tail.}
\label{fig:seq}
\end{figure*}

Comparing the results of the photometric study of the central parts of the dwarf galaxies to the simulations allows us to reconstruct the two main steps of the migration scenario : the infall and merger of star clusters. They are illustrated in Figure \ref{fig:seq}.
We observe a possible migration of massive GCs towards the galaxy center of MATLAS-138 and MATLAS-987. And, given the bright nuclei and the large extent of the tails in MATLAS-207, MATLAS-1216, and MATLAS-1938, we are likely witnessing the merging of two star clusters with a notable mass difference, such as NSC-GC mergers.
Moreover, observations and simulations concur to highlight the rarety of detection of such complex nuclei, as suggested by the computed low detection rate and the short timescales of this formation channel. We also expect to see fewer tidal tails than star clusters in the nuclear region given their low surface brightness and the fact that they are only produced when the merger occurs between the star clusters. Therefore, large surveys of low surface brightness dwarf galaxies with deep and high resolution images are required to observe such processes. We expect to find similar structures in upcoming large space telescope surveys, such as the Euclid Wide Survey\cite{Scaramella2022}.

\newpage
\clearpage

\newpage
\clearpage
\section*{Methods}\label{methods}

\subsection*{MATLAS}
MATLAS\cite{Duc2015,Duc2020} is a deep optical imaging survey exploring the mass assembly of 180 massive early-type galaxies and the build-up of their scaling relations by studying their outermost stellar populations, their fine structures (e.g., tidal tails, shells, and stellar streams), their GC population, as well as their dwarf galaxy satellites. The galaxies are located beyond the Local Volume, at distances between 10 and 45 Mpc, and outside of galaxy cluster environments. The observations were taken with the MegaCam camera of the Canada France Hawaii Telescope (CFHT) between 2012 and 2015, and are composed of 12, 150, 148, 79 images with a $1\deg^2$ field of view in the $u$, $g$, $r$, $i$-band, respectively. The data were reduced with the help of the Elixir-LSB pipeline, providing a surface brightness as deep as $28.5-29$\,mag\,arcsec$^{-2}$ in the $g$-band.

The MATLAS dwarf catalog\cite{Habas2020,Poulain2021} contains 2210 galaxies identified using a visual inspection of the 150 fields in the $g$-band combined with a semi-automated catalog generated using \textsc{source extractor}\cite{Bertin1996} parameters that was visually cleaned to reject potential background sources and to assign a morphological type to each dwarf candidate. The final classification resulted in a sample of 1634 dwarf ellipticals (73\%) and 576 dwarf irregulars (27\%). During the morphological classification, we defined a nucleus as a bright, compact source located within 0.5$R_e$ of the galaxy center, and brighter than any other compact sources within 1$R_e$. This led to a sample of 507 nucleated galaxies. Unless a distance measurement was available, we assumed each dwarf is located at the distance of the central early-type galaxy targeted in the field where the dwarf was detected.

\subsection*{HST follow-up observations}
While the deep MATLAS ground-based images are ideal for detection of low-surface brightness substructure, the average image seeing (0.96" in the \emph{g}-band) limits our ability to resolve the GCs/NSCs and any related structures. Thus, high resolution imaging from space-based observatories, such as the HST, is required. In that context, we conducted follow-up observations in the F606W and F814W filters of the HST ACS camera. The project was designed to characterize the GC population of UDGs\cite{Marleau2024}. Combining the Cycle 27 program GO-16082 (PI: O. Müller) with the Cycle 28 and 29 snapshot programs SNAP-16257 and SNAP-16711 (PI: F. Marleau), we obtained ACS/WFC images for a sample of 79 dwarf galaxies. Given the science goals of the project, a majority of the sample has a low surface brightness and a large size typical of UDGs. Among the targets, we found 41 nucleated dwarfs\cite{Poulain2024}, 13 of which are newly identified in the HST imaging. Serendipitously, we noticed that 10 nucleated dwarfs show a complex nuclear region composed of multiple star clusters and tidal tails.

\subsection*{Dwarf-dwarf merger candidates}
To ensure that our sample is composed of the most likely star cluster merger systems caused by the effect of dynamical friction on GCs alone, excluding any environmental influence on the galaxies, we searched for signs of past galaxy interactions in the light profile of the dwarfs, in particular in their surroundings. Observational signatures of such events include extended tidal tails, stellar shells, or disturbances in the main body of the dwarf. Focusing on the sample with complex nuclei, six of the dwarfs exhibit no visible signs of interactions with nearby galaxies (Extended Data Fig. 1), implying that the nuclear structures most likely formed from internal processes, while four galaxies with asymmetric shapes appear to be tidally perturbed (Extended Data Fig. 2). Some of the latter disturbances could be due to a tidal interaction with their massive host galaxy. Studies\cite{Paudel2014,Paudel2018} reporting tidal interactions between dwarfs and massive galaxies typically find spatial separations between the galaxies within 100\,kpc and the presence of a tidal bridge. For our disturbed objects, the projected distance to the host lies beyond 100\,kpc. Moreover, none of them exhibit tails pointing towards the host. We also investigated the possibility that they could be satellites of another massive galaxy; however, the smallest physical separations -- calculated assuming the dwarf is at the same distance as the potential massive host -- are all greater than 100\,kpc. Therefore, we posit that these disturbances are the result of a dwarf-dwarf interaction or merger. Based on this, we excluded these objects from our sample, leaving us with the six candidates whose complex nuclei most likely formed from an internal process.

\subsection*{Complex nuclei photometry}
We carried out a photometric study of the nuclear substructures in the six non-interacting dwarfs identified in the HST images. The extracted properties of these objects were then compared with those of GCs and NSCs identified in other MATLAS dwarfs\cite{Marleau2024,Poulain2024} and for which photometry could be reliably extracted.
In order to disentangle the different substructures of the complex nuclear regions and measure their photometry, we made use of the software \textsc{mtobjects}\cite{Teeninga2015} (\textsc{mto}). \textsc{mto} is a Max-Tree-based method optimized to detect low-surface brightness sources. We ran the software on each of the complex nuclear regions to produce a segmentation image of the structures. Image segmentation deblends the different sources in an image by grouping the pixels belonging to each objects under the same flag. We produced segmentation images on both the F606W and F814W images to ensure all the structures were successfully detected. For all nuclear regions, we merged the obtained segmentation maps in order to generate a final image including the regions of all the structures. We show the resulting segmentation maps in Extended Data Fig. 3. Using the final segmentation images, we then derived the magnitude and color of each nuclear substructure from the F606W and F814W images. We derived errors on the magnitude by combining the standard error with the standard deviation of the sky. The derived properties are available in Table \ref{tab:photometry}. 
All the nuclear structures have colors in the range of the GCs in the MATLAS dwarfs observed with HST\cite{Marleau2024}, except for one substructure. The substructure is found in a dwarf that exhibits a star cluster with a red color $(m_{F606W}-m_{F814W})_0 = 1.23\pm0.05$ in the typical range of GCs, together with a faint blue stellar tail and clump with $0.03\pm0.05$ and $-0.43\pm0.31$, respectively. Similar to some star-forming nuclei in dwarf elliptical galaxies\cite{Paudel2020} that can be related to a wet-merger scenario. Though an HI study of the MATLAS dwarfs\cite{Poulain2022} reported no HI detection in this galaxy, we opted to remove it from the sample.

\subsection*{N-body simulations}
We performed collisionless N-body simulations of star cluster mergers at the centre of a dwarf galaxy. Our method shares similarities with previous simulations of star cluster mergers in spiral galaxies\cite{Hartmann2011} which we adapted to our HST sample of dwarf galaxies. The set-up of these simulations is designed to model the final stages of the merger. Thus, in their current form, they cannot be used to model the full orbital decay of the star clusters.

We chose to model a dwarf with a cored dark matter (DM) profile using a spherical Burkert halo of mass M$_h = 3\times10^{10}$ M$_{\odot}$ and a scale radius R$_0 = 5.6$ kpc such that the halo radius is R$_h = 3.4$R$_0$\cite{Salucci2000}. The dwarf also has a double exponential stellar disk of mass $log(M_{\star}/M_{\odot}) = 7.7$, $R_e = 1.5$ kpc and an axial ratio of 0.6 (i.e., a thick disk), corresponding to the median $M_{\star}$ and $R_e$ of the MATLAS dwarfs observed by HST. In this way, the merging clusters also interact gravitationally with both the dark matter halo and stellar disk of the model dwarf galaxy. The halo consists of $10^7$ DM particles of mass 300 M$_{\odot}$, as well as just over $10^5$ stellar particles of mass 500 M$_{\odot}$. In addition, we tested the effect of a change in the stellar component with a 10 times more massive stellar disk. We also modified the DM profile by switching from a cored Burkert to a cuspy NFW halo of equal halo mass and NFW concentration c=11, typical for this mass\cite{DiemerJoyce2019}.

We simulated three types of cluster mergers: NSC-NSC, NSC-GC, and GC-GC. We use a Sérsic model for the NSC and GC. We set the Sersic parameters of the NSC profile according to the average properties of our HST sample\cite{Poulain2024}, such that $log(M_{\star}/M_{\odot}) = 6.5$, $R_e = 7$ pc, and Sérsic index $n = 2$. The Sersic parameters of the GC are based on the average parameters of the GCs in Andromeda and the Milky Way. That is $log(M_{\star}/M_{\odot}) = 5.2$, $R_e = 3.2$ pc, and $n = 1$. These Sersic models consist of $5\times10^5$ stars of mass 6.3 M$_{\odot}$ and $10^5$ stars of mass 1.6 M$_{\odot}$ for the NSC and GC, respectively. In the case of a NSC-NSC merger or GC-GC merger, the merger mass ratio is 1:1. We also test the case of multiple sequential GC mergers until up to 5 GCs have been merged into a single remnant. For NSC-GC mergers, we test mass ratios of 5, 10 and 20.

The simulations were conducted using the adaptive mesh refinement code \textsc{Ramses}\cite{Teyssier2002}. The simulation volume has a side length of 200 kpc which encompasses the entire dark matter halo of the model galaxy. We choose a refinement grid where the maximum level of refinement varies with distance from the dwarf galaxy centre. Inside of 0.8 kpc, which corresponds to the region in which the clusters merge and interact with the dwarf galaxy mass distribution (the 'merger region'), a resolution of 0.375 pc is reached. Outside of the merger region, the resolution steadily steps down (at radii of 2, 10, 20, 40, 80, and 160 kpc) reaching a minimum resolution of 6.25 kpc in the far outskirts of the DM halo. In this way, we can model the dynamical evolution of the entire galaxy, while restricting the region of maximum resolution to the area where it is required. We also conducted several simulations at a lower resolution of 1.5 pc. We find this resolution change does not affect our main conclusions.

The simulations are performed over time durations ranging from 75 to 278\,Myr, depending on the time required for the merger process to be fully completed. Snapshots were produced frequently, with an output every 3.73\,Myr in order to capture the rapidly changing details of the merger process. Typically, a star cluster (either NSC or GC) collides with another cluster located at the center of the dwarf. We set the initial position of the infalling star cluster to be 200\,pc from the galaxy center. A fiducial radial velocity of 2.5\,km/s and tangential velocity of 1.5\,km/s is chosen. However, we also tested the impact of increasing the radial velocity up to 5, 10, 25 and 50\,km/s, and varying the tangential velocity from 0 to 0.5, 1.5, 2, 5, and 7.5\,km/s, to look for possible impacts on the structures formed during the merger process. The tangential velocity of 5\,km/s is similar to the circular velocity of the dwarf galaxy model at the radius where the star clusters are interacting. Overall, as long as the two clusters end up colliding, we find the appearance and duration of these structures are not strongly affected by these choices. The two exceptions are the highest radial and tangential velocity considered, for which no merger happens within 0.5\,Gyr and flybys occur without creating any tidal features.

\backmatter

\bmhead{Acknowledgments}
This version of the article has been accepted for publication, after peer review but is not the Version of Record and does not reflect post-acceptance improvements, or any corrections. The Version of Record is available online at
\url{https://doi.org/10.1038/s41586-025-08783-9}.
This research is based on observations from the NASA/ESA Hubble Space Telescope obtained at the Space Telescope Science Institute, which is operated by the Association of Universities for Research in Astronomy, Incorporated, under NASA contract NAS5-26555. Support for Program number GO-16257 and GO-16711 was provided through a grant from the STScI under NASA contract NAS5-26555. This research was supported by the International Space Science Institute (ISSI) in Bern, through ISSI International Team project No. 534. M.P. is supported by the Academy of Finland grant No. 347089. O.M. is grateful to the Swiss National Science Foundation for financial support under the grant number PZ00P2\_202104. P.R.D. gratefully acknowledges support from grant HST-GO-16257.002-A. S.P. acknowledges support from the Mid-career Researcher Program (No. RS-2023-00208957). S.L. acknowledges the support from the Sejong Science Fellowship Program by the National Research Foundation of Korea (NRF) grant funded by the Korea government (MSIT) (No. NRF-2021R1C1C2006790). RH acknowledges funding from the Italian INAF Large Grant 12 - 2022. RS acknowledges financial support from FONDECYT Regular 2023 project No. 1230441 and also gratefully acknowledges financial support from ANID - MILENIO - NCN2024\_112.

\bmhead{Author contributions}
M.P., F.R.M., and P.A.D. designed this study.
M.P. led and performed the data analysis, and wrote the manuscript.
R.S. performed the N-body simulations.
F.R.M. and O.M. are the PIs of the HST observing programs.
R.H., P.R.D, J.F, S.L, S.P, and R.S.J. provided significant feedback on the work.
 
%All authors contributed to the writing of the manuscript by editing or providing feedback.

\bmhead{Competing interests}
The authors declare no competing interests.

\bmhead{Data availability}
The Hubble Space Telescope data used in this analysis are available at the Mikulski Archive for Space Telescopes (MAST) with the programme IDs GO-16082, SNAP-16257, and SNAP-16711.

\bmhead{Code availability}
This work made use of the software \textsc{mtobjects}, and \textsc{python3} widely used packages such as Matplotlib, NumPy, SciPy, and Astropy. The simulations are based on \textsc{dice} for galaxy model set up, \textsc{ramses} to run it, and \textsc{rdramses} for analysis. All codes and software are publicly available.

\bmhead{Correspondence and requests for materials} should be addressed to Mélina Poulain.

\bmhead{Reprints and permissions information} is available at http://www.nature.com/reprints.

%\bibliography{sn-bibliography}

\begin{thebibliography}{10}
\expandafter\ifx\csname url\endcsname\relax
  \def\url#1{\burl{#1}}\fi
\expandafter\ifx\csname urlprefix\endcsname\relax\def\urlprefix{URL }\fi
\providecommand{\bibinfo}[2]{#2}
\providecommand{\eprint}[2][]{\url{#2}}
\providecommand{\doi}[1]{\url{https://doi.org/#1}}
\bibcommenthead

\bibitem{Cote2006}
\bibinfo{author}{C{\^{o}}t{\'{e}}, P.} \emph{et~al.}
\newblock \bibinfo{title}{The {ACS} virgo cluster survey. {VIII}. the nuclei of early-type galaxies}.
\newblock \emph{\bibinfo{journal}{ApJs}} \textbf{\bibinfo{volume}{165}}, \bibinfo{pages}{57--94} (\bibinfo{year}{2006}).

\bibitem{Janssen2019}
\bibinfo{author}{{S{\'a}nchez-Janssen}, R.} \emph{et~al.}
\newblock \bibinfo{title}{{The Next Generation Virgo Cluster Survey. XXIII. Fundamentals of Nuclear Star Clusters over Seven Decades in Galaxy Mass}}.
\newblock \emph{\bibinfo{journal}{ApJ}} \textbf{\bibinfo{volume}{878}}, \bibinfo{pages}{18} (\bibinfo{year}{2019}).

\bibitem{Loose1982}
\bibinfo{author}{{Loose}, H.~H.}, \bibinfo{author}{{Kruegel}, E.} \& \bibinfo{author}{{Tutukov}, A.}
\newblock \bibinfo{title}{{Bursts of star formation in the galactic centre}}.
\newblock \emph{\bibinfo{journal}{A\&A}} \textbf{\bibinfo{volume}{105}}, \bibinfo{pages}{342--350} (\bibinfo{year}{1982}).

\bibitem{Tremaine1975}
\bibinfo{author}{{Tremaine}, S.~D.}, \bibinfo{author}{{Ostriker}, J.~P.} \& \bibinfo{author}{{Spitzer}, J., L.}
\newblock \bibinfo{title}{{The formation of the nuclei of galaxies. I. M31.}}
\newblock \emph{\bibinfo{journal}{ApJ}} \textbf{\bibinfo{volume}{196}}, \bibinfo{pages}{407--411} (\bibinfo{year}{1975}).

\bibitem{Turner2012}
\bibinfo{author}{{Turner}, M.~L.} \emph{et~al.}
\newblock \bibinfo{title}{{The ACS Fornax Cluster Survey. VI. The Nuclei of Early-type Galaxies in the Fornax Cluster}}.
\newblock \emph{\bibinfo{journal}{ApJs}} \textbf{\bibinfo{volume}{203}}, \bibinfo{pages}{5} (\bibinfo{year}{2012}).

\bibitem{Johnston2020}
\bibinfo{author}{{Johnston}, E.~J.} \emph{et~al.}
\newblock \bibinfo{title}{{The Next Generation Fornax Survey (NGFS): VII. A MUSE view of the nuclear star clusters in Fornax dwarf galaxies}}.
\newblock \emph{\bibinfo{journal}{MNRAS}} \textbf{\bibinfo{volume}{495}}, \bibinfo{pages}{2247--2264} (\bibinfo{year}{2020}).

\bibitem{Fahrion2021}
\bibinfo{author}{{Fahrion}, K.} \emph{et~al.}
\newblock \bibinfo{title}{{Diversity of nuclear star cluster formation mechanisms revealed by their star formation histories}}.
\newblock \emph{\bibinfo{journal}{A\&A}} \textbf{\bibinfo{volume}{650}}, \bibinfo{pages}{A137} (\bibinfo{year}{2021}).

\bibitem{Fahrion2022}
\bibinfo{author}{{Fahrion}, K.}, \bibinfo{author}{{Leaman}, R.}, \bibinfo{author}{{Lyubenova}, M.} \& \bibinfo{author}{{van de Ven}, G.}
\newblock \bibinfo{title}{{Disentangling the formation mechanisms of nuclear star clusters}}.
\newblock \emph{\bibinfo{journal}{A\&A}} \textbf{\bibinfo{volume}{658}}, \bibinfo{pages}{A172} (\bibinfo{year}{2022}).

\bibitem{Fahrion2022b}
\bibinfo{author}{{Fahrion}, K.} \emph{et~al.}
\newblock \bibinfo{title}{{Nuclear star cluster formation in star-forming dwarf galaxies}}.
\newblock \emph{\bibinfo{journal}{A\&A}} \textbf{\bibinfo{volume}{667}}, \bibinfo{pages}{A101} (\bibinfo{year}{2022}).

\bibitem{Neumayer2020}
\bibinfo{author}{{Neumayer}, N.}, \bibinfo{author}{{Seth}, A.} \& \bibinfo{author}{{B{\"o}ker}, T.}
\newblock \bibinfo{title}{{Nuclear star clusters}}.
\newblock \emph{\bibinfo{journal}{A\&Ar}} \textbf{\bibinfo{volume}{28}}, \bibinfo{pages}{4} (\bibinfo{year}{2020}).

\bibitem{Lauer1993}
\bibinfo{author}{{Lauer}, T.~R.} \emph{et~al.}
\newblock \bibinfo{title}{{Planetary Camera Observations of the Double Nucleus of M31}}.
\newblock \emph{\bibinfo{journal}{AJ}} \textbf{\bibinfo{volume}{106}}, \bibinfo{pages}{1436} (\bibinfo{year}{1993}).

\bibitem{Lauer1996}
\bibinfo{author}{{Lauer}, T.~R.} \emph{et~al.}
\newblock \bibinfo{title}{{Hubble Space Telescope Observations of the Double Nucleus of NGC 4486B}}.
\newblock \emph{\bibinfo{journal}{ApJ}} \textbf{\bibinfo{volume}{471}}, \bibinfo{pages}{L79} (\bibinfo{year}{1996}).

\bibitem{Debattista2006}
\bibinfo{author}{{Debattista}, V.~P.} \emph{et~al.}
\newblock \bibinfo{title}{{The Binary Nucleus in VCC 128: A Candidate Supermassive Black Hole in a Dwarf Elliptical Galaxy}}.
\newblock \emph{\bibinfo{journal}{ApJ}} \textbf{\bibinfo{volume}{651}}, \bibinfo{pages}{L97--L100} (\bibinfo{year}{2006}).

\bibitem{Menezes2018}
\bibinfo{author}{{Menezes}, R.~B.} \& \bibinfo{author}{{Steiner}, J.~E.}
\newblock \bibinfo{title}{{Double Nuclei in NGC 908 and NGC 1187}}.
\newblock \emph{\bibinfo{journal}{ApJ}} \textbf{\bibinfo{volume}{868}}, \bibinfo{pages}{67} (\bibinfo{year}{2018}).

\bibitem{Georgiev2014}
\bibinfo{author}{{Georgiev}, I.~Y.} \& \bibinfo{author}{{B{\"o}ker}, T.}
\newblock \bibinfo{title}{{Nuclear star clusters in 228 spiral galaxies in the HST/WFPC2 archive: catalogue and comparison to other stellar systems}}.
\newblock \emph{\bibinfo{journal}{MNRAS}} \textbf{\bibinfo{volume}{441}}, \bibinfo{pages}{3570--3590} (\bibinfo{year}{2014}).

\bibitem{Schiavi2021}
\bibinfo{author}{{Schiavi}, R.}, \bibinfo{author}{{Capuzzo-Dolcetta}, R.}, \bibinfo{author}{{Georgiev}, I.~Y.}, \bibinfo{author}{{Arca-Sedda}, M.} \& \bibinfo{author}{{Mastrobuono-Battisti}, A.}
\newblock \bibinfo{title}{{Are we observing an NSC in course of formation in the NGC 4654 galaxy?}}
\newblock \emph{\bibinfo{journal}{MNRAS}} \textbf{\bibinfo{volume}{503}}, \bibinfo{pages}{594--602} (\bibinfo{year}{2021}).

\bibitem{Fahrion2024}
\bibinfo{author}{{Fahrion}, K.} \emph{et~al.}
\newblock \bibinfo{title}{{Growing a nuclear star cluster from star formation and cluster mergers: The JWST NIRSpec view of NGC 4654}}.
\newblock \emph{\bibinfo{journal}{A\&A}} \textbf{\bibinfo{volume}{687}}, \bibinfo{pages}{A83} (\bibinfo{year}{2024}).

\bibitem{Guillard2016}
\bibinfo{author}{{Guillard}, N.}, \bibinfo{author}{{Emsellem}, E.} \& \bibinfo{author}{{Renaud}, F.}
\newblock \bibinfo{title}{{New insights on the formation of nuclear star clusters}}.
\newblock \emph{\bibinfo{journal}{MNRAS}} \textbf{\bibinfo{volume}{461}}, \bibinfo{pages}{3620--3629} (\bibinfo{year}{2016}).

\bibitem{Pak2016}
\bibinfo{author}{{Pak}, M.}, \bibinfo{author}{{Paudel}, S.}, \bibinfo{author}{{Lee}, Y.} \& \bibinfo{author}{{Kim}, S.~C.}
\newblock \bibinfo{title}{{MCG+08-22-082: A Double Core and Boxy Appearance Dwarf Lenticular Galaxy Suspected to be a Merger Remnant}}.
\newblock \emph{\bibinfo{journal}{AJ}} \textbf{\bibinfo{volume}{151}}, \bibinfo{pages}{141} (\bibinfo{year}{2016}).

\bibitem{Marleau2024}
\bibinfo{author}{{Marleau}, F.~R.} \emph{et~al.}
\newblock \bibinfo{title}{{Dwarf galaxies in the MATLAS Survey: Hubble Space Telescope observations of the globular cluster systems of 74 ultra-diffuse galaxies}}.
\newblock \emph{\bibinfo{journal}{A\&A}} \textbf{\bibinfo{volume}{690}}, \bibinfo{pages}{A339} (\bibinfo{year}{2024}).

\bibitem{Marleau2021}
\bibinfo{author}{{Marleau}, F.~R.} \emph{et~al.}
\newblock \bibinfo{title}{{Ultra diffuse galaxies in the MATLAS low-to-moderate density fields}}.
\newblock \emph{\bibinfo{journal}{A\&A}} \textbf{\bibinfo{volume}{654}}, \bibinfo{pages}{A105} (\bibinfo{year}{2021}).

\bibitem{Poulain2024}
\bibinfo{author}{{Poulain}, M.} \emph{et~al.}
\newblock \bibinfo{title}{Dwarf galaxies in the matlas survey: Hubble space telescope observations of nuclear star clusters}.
\newblock \bibinfo{note}{In preparation}.

\bibitem{Harris1996}
\bibinfo{author}{{Harris}, W.~E.}
\newblock \bibinfo{title}{{A Catalog of Parameters for Globular Clusters in the Milky Way}}.
\newblock \emph{\bibinfo{journal}{A\&A}} \textbf{\bibinfo{volume}{112}}, \bibinfo{pages}{1487} (\bibinfo{year}{1996}).

\bibitem{Peacock2010}
\bibinfo{author}{{Peacock}, M.~B.} \emph{et~al.}
\newblock \bibinfo{title}{{The M31 globular cluster system: ugriz and K-band photometry and structural parameters}}.
\newblock \emph{\bibinfo{journal}{MNRAS}} \textbf{\bibinfo{volume}{402}}, \bibinfo{pages}{803--818} (\bibinfo{year}{2010}).

\bibitem{Scaramella2022}
\bibinfo{author}{{Euclid Collaboration: Scaramella}, R.} \emph{et~al.}
\newblock \bibinfo{title}{{Euclid preparation. I. The Euclid Wide Survey}}.
\newblock \emph{\bibinfo{journal}{A\&A}} \textbf{\bibinfo{volume}{662}}, \bibinfo{pages}{A112} (\bibinfo{year}{2022}).
\end{thebibliography}

\begin{thebibliography}{10}
\expandafter\ifx\csname url\endcsname\relax
  \def\url#1{\burl{#1}}\fi
\expandafter\ifx\csname urlprefix\endcsname\relax\def\urlprefix{URL }\fi
\providecommand{\bibinfo}[2]{#2}
\providecommand{\eprint}[2][]{\url{#2}}
\providecommand{\doi}[1]{\url{https://doi.org/#1}}
\bibcommenthead
\setcounter{NAT@ctr}{25}

\bibitem{Duc2015}
\bibinfo{author}{{Duc}, P.-A.} \emph{et~al.}
\newblock \bibinfo{title}{{The ATLAS$^{3D}$ project - XXIX. The new look of early-type galaxies and surrounding fields disclosed by extremely deep optical images}}.
\newblock \emph{\bibinfo{journal}{MNRAS}} \textbf{\bibinfo{volume}{446}}, \bibinfo{pages}{120--143} (\bibinfo{year}{2015}).

\bibitem{Duc2020}
\bibinfo{author}{{Duc}, P.-A.}
\newblock \bibinfo{title}{{MATLAS: a deep exploration of the surroundings of massive early-type galaxies}}.
\newblock \emph{\bibinfo{journal}{arXiv e-prints}} \bibinfo{pages}{arXiv:2007.13874} (\bibinfo{year}{2020}).

\bibitem{Habas2020}
\bibinfo{author}{{Habas}, R.} \emph{et~al.}
\newblock \bibinfo{title}{{Newly discovered dwarf galaxies in the MATLAS low-density fields}}.
\newblock \emph{\bibinfo{journal}{MNRAS}} \textbf{\bibinfo{volume}{491}}, \bibinfo{pages}{1901--1919} (\bibinfo{year}{2020}).

\bibitem{Poulain2021}
\bibinfo{author}{{Poulain}, M.} \emph{et~al.}
\newblock \bibinfo{title}{{Structure and morphology of the MATLAS dwarf galaxies and their central nuclei}}.
\newblock \emph{\bibinfo{journal}{MNRAS}} \textbf{\bibinfo{volume}{506}}, \bibinfo{pages}{5494--5511} (\bibinfo{year}{2021}).

\bibitem{Bertin1996}
\bibinfo{author}{{Bertin}, E.} \& \bibinfo{author}{{Arnouts}, S.}
\newblock \bibinfo{title}{{SExtractor: Software for source extraction.}}
\newblock \emph{\bibinfo{journal}{A\&AS}} \textbf{\bibinfo{volume}{117}}, \bibinfo{pages}{393--404} (\bibinfo{year}{1996}).

\bibitem{Paudel2014}
\bibinfo{author}{{Paudel}, S.} \& \bibinfo{author}{{Ree}, C.~H.}
\newblock \bibinfo{title}{{Tidal Interaction as the Origin of Early-type Dwarf Galaxies in Group Environments}}.
\newblock \emph{\bibinfo{journal}{ApJL}} \textbf{\bibinfo{volume}{796}}, \bibinfo{pages}{L14} (\bibinfo{year}{2014}).

\bibitem{Paudel2018}
\bibinfo{author}{{Paudel}, S.}, \bibinfo{author}{{Smith}, R.}, \bibinfo{author}{{Yoon}, S.~J.}, \bibinfo{author}{{Calder{\'o}n-Castillo}, P.} \& \bibinfo{author}{{Duc}, P.-A.}
\newblock \bibinfo{title}{{A Catalog of Merging Dwarf Galaxies in the Local Universe}}.
\newblock \emph{\bibinfo{journal}{ApJS}} \textbf{\bibinfo{volume}{237}}, \bibinfo{pages}{36} (\bibinfo{year}{2018}).

\bibitem{Teeninga2015}
\bibinfo{author}{Teeninga, P.}, \bibinfo{author}{Moschini, U.}, \bibinfo{author}{Trager, S.~C.} \& \bibinfo{author}{Wilkinson, M. H.~F.}
\newblock \bibinfo{editor}{Benediktsson, J.~A.}, \bibinfo{editor}{Chanussot, J.}, \bibinfo{editor}{Najman, L.} \& \bibinfo{editor}{Talbot, H.} (eds) \emph{\bibinfo{title}{Improved detection of faint extended astronomical objects through statistical attribute filtering}}.
\newblock (eds \bibinfo{editor}{Benediktsson, J.~A.}, \bibinfo{editor}{Chanussot, J.}, \bibinfo{editor}{Najman, L.} \& \bibinfo{editor}{Talbot, H.}) \emph{\bibinfo{booktitle}{Mathematical Morphology and Its Applications to Signal and Image Processing}}, \bibinfo{pages}{157--168} (\bibinfo{publisher}{Springer International Publishing}, \bibinfo{address}{Cham}, \bibinfo{year}{2015}).

\bibitem{Paudel2020}
\bibinfo{author}{{Paudel}, S.} \& \bibinfo{author}{{Yoon}, S.-J.}
\newblock \bibinfo{title}{{Starbursting Nuclei in Old Dwarf Galaxies}}.
\newblock \emph{\bibinfo{journal}{ApJL}} \textbf{\bibinfo{volume}{898}}, \bibinfo{pages}{L47} (\bibinfo{year}{2020}).

\bibitem{Poulain2022}
\bibinfo{author}{{Poulain}, M.} \emph{et~al.}
\newblock \bibinfo{title}{{HI observations of the MATLAS dwarf and ultra-diffuse galaxies}}.
\newblock \emph{\bibinfo{journal}{A\&A}} \textbf{\bibinfo{volume}{659}}, \bibinfo{pages}{A14} (\bibinfo{year}{2022}).

\bibitem{Hartmann2011}
\bibinfo{author}{{Hartmann}, M.}, \bibinfo{author}{{Debattista}, V.~P.}, \bibinfo{author}{{Seth}, A.}, \bibinfo{author}{{Cappellari}, M.} \& \bibinfo{author}{{Quinn}, T.~R.}
\newblock \bibinfo{title}{{Constraining the role of star cluster mergers in nuclear cluster formation: simulations confront integral-field data}}.
\newblock \emph{\bibinfo{journal}{MNRAS}} \textbf{\bibinfo{volume}{418}}, \bibinfo{pages}{2697--2714} (\bibinfo{year}{2011}).

\bibitem{Salucci2000}
\bibinfo{author}{{Salucci}, P.} \& \bibinfo{author}{{Burkert}, A.}
\newblock \bibinfo{title}{{Dark Matter Scaling Relations}}.
\newblock \emph{\bibinfo{journal}{ApJL}} \textbf{\bibinfo{volume}{537}}, \bibinfo{pages}{L9--L12} (\bibinfo{year}{2000}).

\bibitem{DiemerJoyce2019}
\bibinfo{author}{{Diemer}, B.} \& \bibinfo{author}{{Joyce}, M.}
\newblock \bibinfo{title}{{An Accurate Physical Model for Halo Concentrations}}.
\newblock \emph{\bibinfo{journal}{ApJ}} \textbf{\bibinfo{volume}{871}}, \bibinfo{pages}{168} (\bibinfo{year}{2019}).

\bibitem{Teyssier2002}
\bibinfo{author}{{Teyssier}, R.}
\newblock \bibinfo{title}{{Cosmological hydrodynamics with adaptive mesh refinement. A new high resolution code called RAMSES}}.
\newblock \emph{\bibinfo{journal}{A\&A}} \textbf{\bibinfo{volume}{385}}, \bibinfo{pages}{337--364} (\bibinfo{year}{2002}).

\end{thebibliography}

\begin{appendices}
%\section{Extended Data}

\renewcommand{\figurename}{} % redefine the command that creates the figure name
\renewcommand\thefigure{Extended Data Fig. \arabic{figure}}

\begin{figure*}
\centering
%\begin{subfigure}{\textwidth}
%\centering
\includegraphics[width=0.8\linewidth]{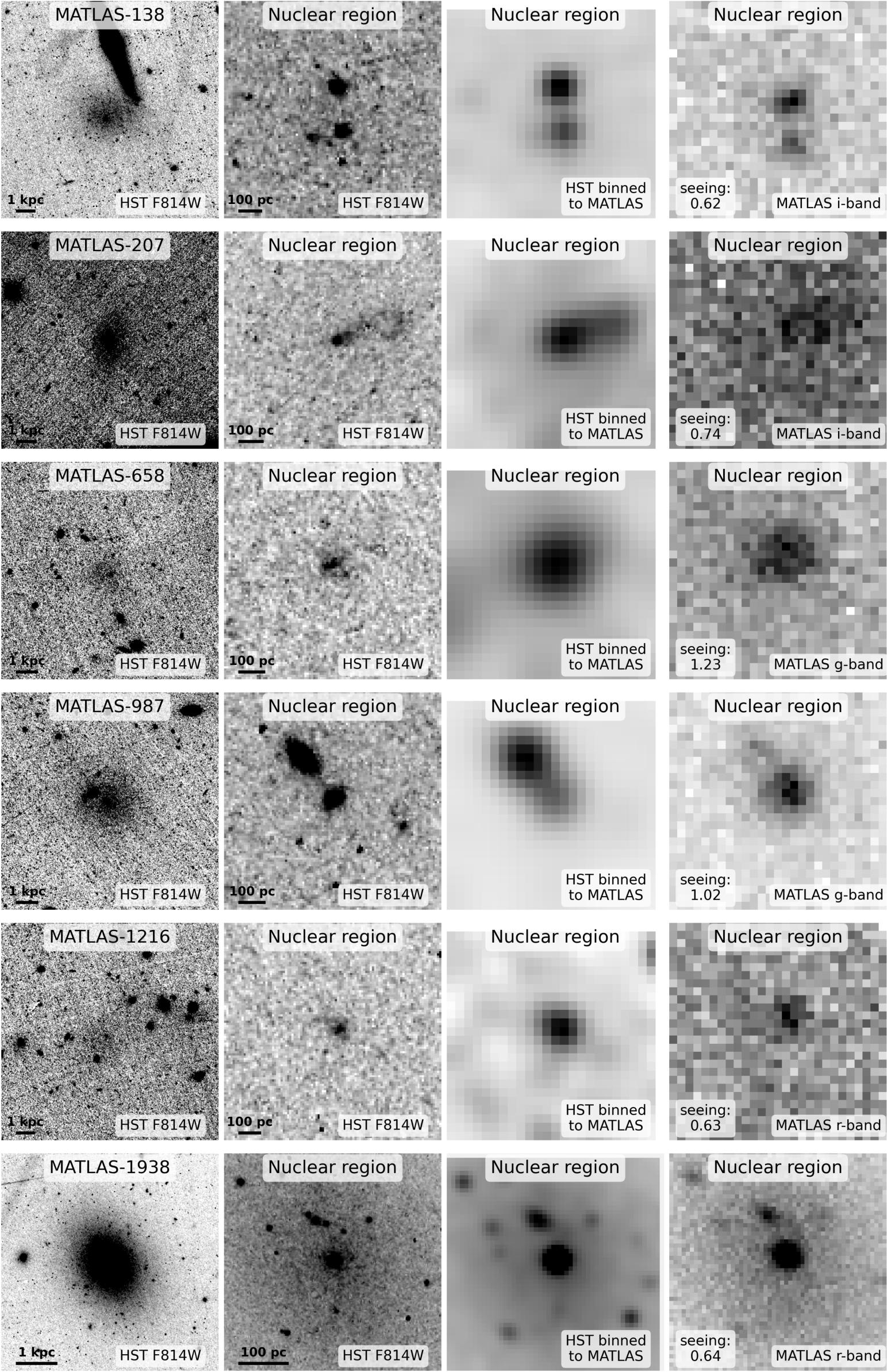}
%\caption{The six non-interacting dwarfs}
%\end{subfigure}
\caption{\footnotesize \textbf{The six non-interacting dwarf galaxies showing a complex nucleus.} For each galaxy, \emph{from left to right}: 1'$\times$1' cutout of the dwarf in the HST F814W filter, the central 5"$\times$5" region of the dwarf, the nuclear region cutout 4pix$\times$4pix binned to which we applied a Gaussian filter to obtain a seeing similar to MATLAS, and the corresponding MATLAS observation in the band with the best seeing. MATLAS-1938 has a 10"$\times$10" zoom into its nuclear region. All images have North up and East left.}
%\label{fig:noninteract_dwarfs}
%\label{fig:dwarfs_HST_MATLAS}
\end{figure*}

\begin{figure*}
%\ContinuedFloat
\centering
%\begin{subfigure}{\textwidth}
%\centering
\includegraphics[width=0.8\linewidth]{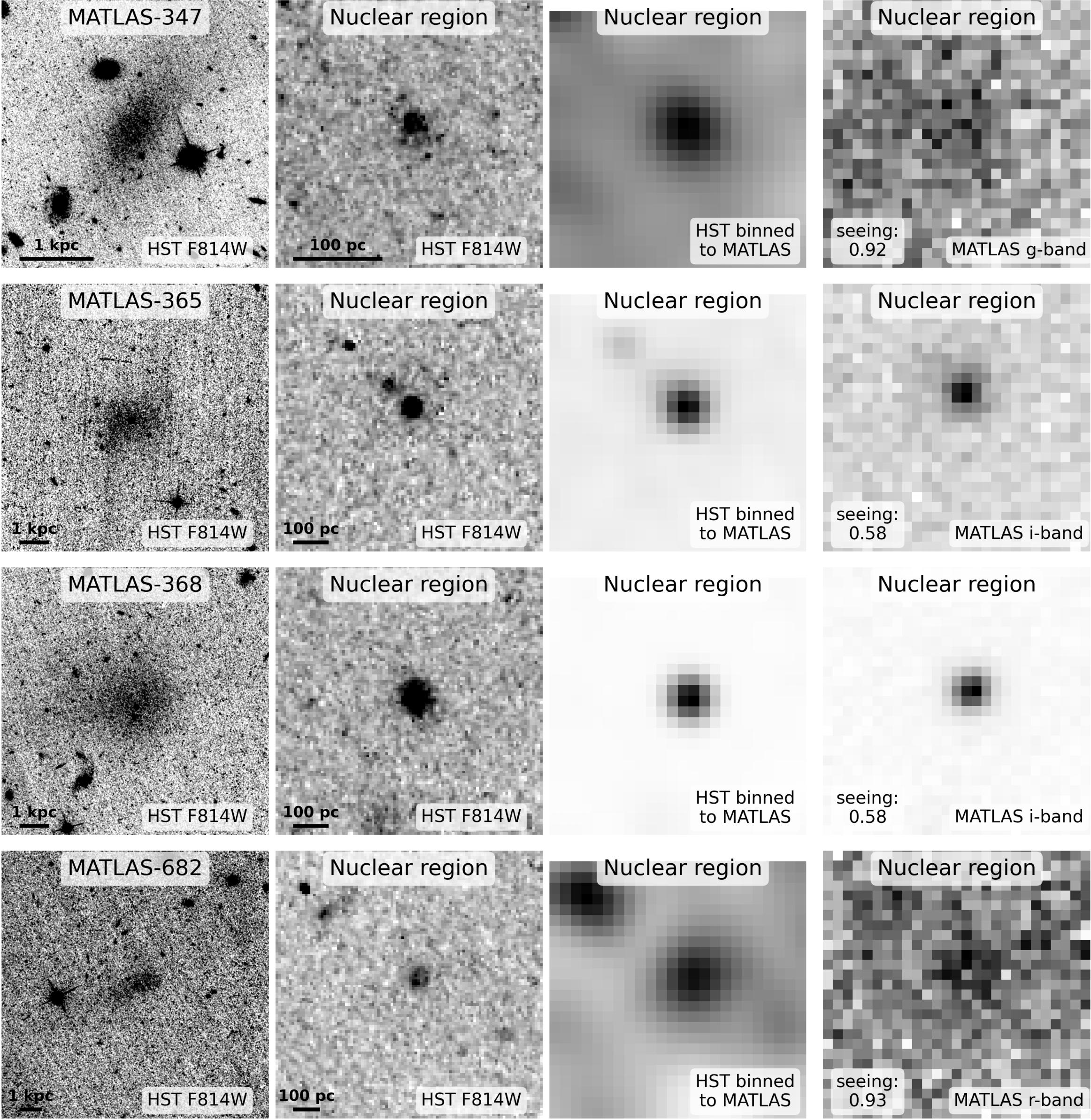}
%\caption{The four dwarf-dwarf merger candidates}
%\end{subfigure}
\caption{\footnotesize \textbf{The four dwarf-dwarf merger candidates showing a complex nucleus.} For each galaxy, \emph{from left to right}: 1'$\times$1' cutout of the dwarf in the HST F814W filter, the central 5"$\times$5" region of the dwarf, the nuclear region cutout 4pix$\times$4pix binned to which we applied a Gaussian filter to obtain a seeing similar to MATLAS, and the corresponding MATLAS observation in the band with the best seeing. All images have North up and East left.}%MATLAS-1938 has a 10"$\times$10" zoom into its nuclear region.}
%\label{fig:disturbed_dwarfs}
\end{figure*}

\begin{figure*}
\centering
\includegraphics[width=\linewidth]{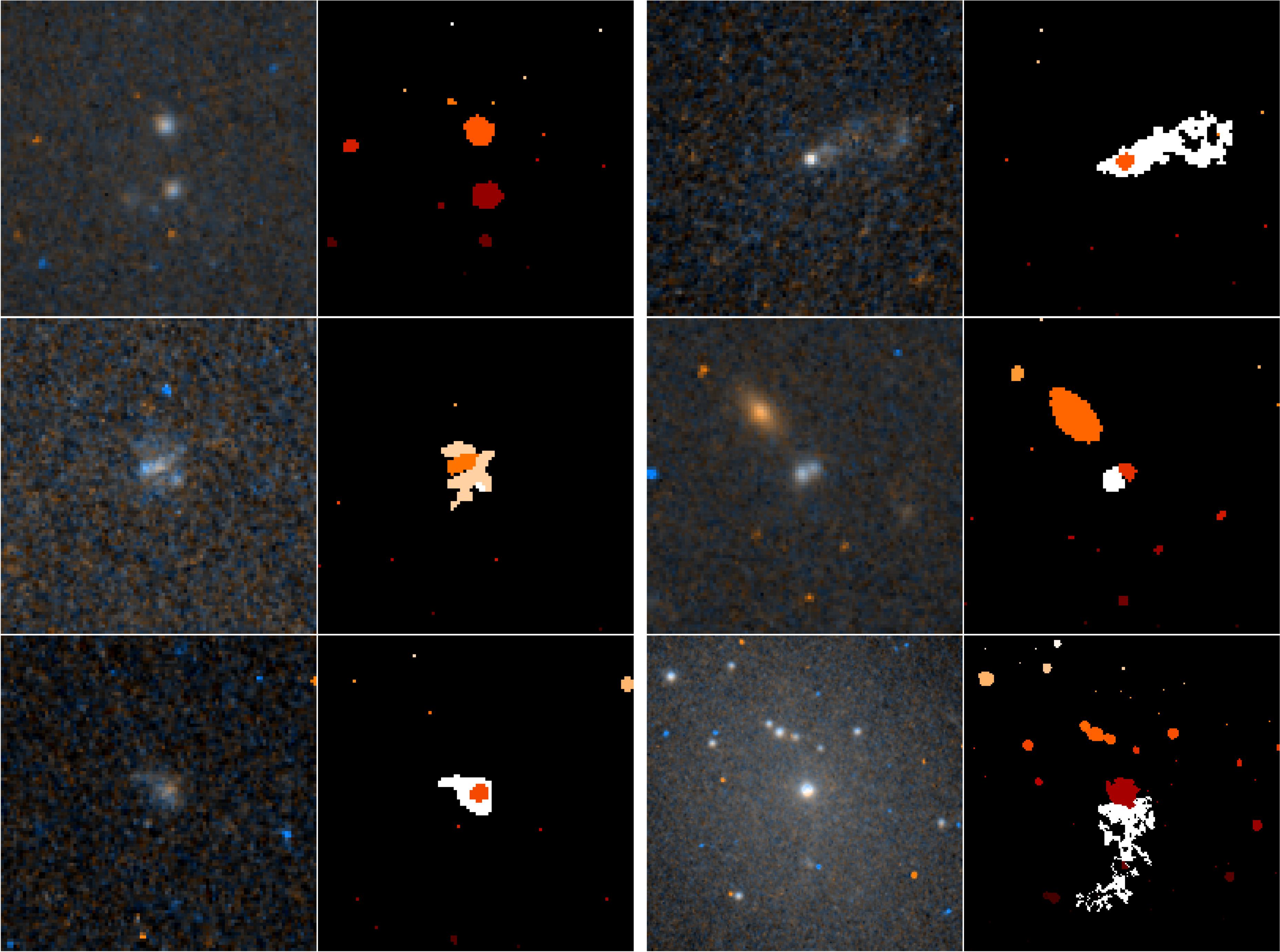}
\caption{\footnotesize \textbf{Composed segmentation images from \textsc{MTO} displaying the detection of the different structures within the nuclear region of the dwarfs.} \emph{From top to bottom}, \emph{left to right}: MATLAS-138, MATLAS-207, MATLAS-658, MATLAS-987, MATLAS-1216, MATLAS-1938.}
%\label{fig:mto}
\end{figure*}

\newpage
\clearpage
\renewcommand{\tablename}{Extended Data Table}
\begin{table}

   \caption{\label{tab:photometry}\textbf{Derived photometric properties.}}
   \begin{tabular}{ccccccc}
    \toprule
     Dwarf ID & RA & Dec & Dist & Structure & $m_{F606W}$ & $(m_{F606W}-m_{F814W})_0$\\
      & [deg] & [deg] & [Mpc] &  & [mag] & [mag] \\
    \toprule
    MATLAS-138 & 27.8987 & 22.294 & 37.5 & cluster & $23.26\pm0.02$ & $0.75\pm0.03$\\
     &  &  &  & cluster & $23.86\pm0.02$ & $0.81\pm0.03$\\
    MATLAS-207 & 42.4458 & -1.1774 & 35.3 & cluster & $25.40\pm0.04$ & $0.90\pm0.05$\\
     &  &  &  & tail & $24.53\pm0.04$ & $1.02\pm0.05$\\
    MATLAS-658 & 154.4670 & 22.3337 & 33.1 & cluster & $25.67\pm0.04$ & $1.23\pm0.05$\\
     &  &  &  & tail & $24.31\pm0.02$ & $0.03\pm0.05$\\
     &  &  &  & cluster & $27.59\pm0.11$ & $-0.43\pm0.31$\\
    MATLAS-987 & 170.6945 & 38.4596 & 32.7 & cluster & $24.22\pm0.03$ & $0.63\pm0.04$\\
     &  &  &  & cluster & $23.47\pm0.03$ & $0.70\pm0.04$\\
    MATLAS-1216 & 183.7925 & 7.3112 & 39.2 & cluster & $26.10\pm0.05$ & $1.08\pm0.06$\\
     &  &  &  & tail & $25.11\pm0.03$ & $0.54\pm0.06$\\
    MATLAS-1938 & 225.1376 & 2.2304 & 17.8* & cluster & $21.04\pm0.01$ & $0.74\pm0.01$\\
     &  &  &  & tail & $23.89\pm0.04$ & $0.90\pm0.05$\\
    \bottomrule
		\end{tabular}
   \begin{tablenotes}
      \item * Distance measurement from SDSS DR13 database.
      \item The magnitudes and colors are extinction corrected, using the NASA/IPAC Extragalactic Database (NED) Galactic Extinction Calculator.
    \end{tablenotes}
\end{table}

\end{appendices}

\end{document}